\documentclass[11pt,twoside]{article} 
\usepackage{asp2004}
\usepackage{epsf}
\usepackage{psfig}
\usepackage{lscape} 

\begin{document} 

\title{White Dwarfs in Open Clusters: Calibrating the Clock}

\author{T. von Hippel, D.E. Winget, W.H. Jefferys, and J.G. Scott}

\affil{The University of Texas at Austin, 1 University Station C1400,
Austin, TX 78712, USA}

\begin{abstract} 
We present an update of our on-going effort to improve the precision of
white dwarf cosmochronology via careful analyses of white dwarf photometry
in open clusters.  To improve the precision of white dwarf and main
sequence age analysis, we are developing a new interpretative scheme using
a Bayesian statistical approach that matches observations to simulated
stellar clusters.  Here we present our first tests of the Bayesian
approach with simulated stellar clusters with ages of 1, 2, and 4 billion
years.
\end{abstract}

\section{Introduction}

The goal of our open cluster observations and modeling is to improve the
precision of white dwarf cosmochronology by comparing white dwarf (WD)
cooling ages to main sequence turn-off (MSTO) ages in open clusters
spanning a wide range of age and metallicity.  Cluster observations to
date demonstrate a good overall agreement between WD and MSTO ages for
clusters spanning the age range of $\sim$160 Myr to $\sim$4 Gyr (von
Hippel 2005).  Our future work will focus on pushing the well observed
cluster sample to greater ages, to a wider metallicity range, and to
increasing the precision for the cluster WD and MSTO age determinations.
Here we focus on our approach to increasing the precision of the age
determinations.

\section{New Computational Approach}

We apply a new modeling plus Bayesian technique to determine precise white
dwarf cooling ages and standard stellar evolutionary ages.  For the
examples here, the ingredients are a Miller \& Scalo (1979) Initial Mass
Function, Girardi et al. (2000) stellar evolution time scales, the
Initial-Final Mass Relation of Weidemann (2000), WD cooling time scales of
Wood (1992), and the WD atmosphere colors of Bergeron et al. (1995).
Other variants on these modeling ingredients will be incorporated in the
future.  Our Bayesian approach uses a Markov Chain Monte Carlo scheme to
recover the distribution in the parameters of interest, which include
masses for every cluster star, as well as the cluster-wide properties of
age, distance, metallicity, and reddening.

\begin{figure}[!ht]
\plotfiddle{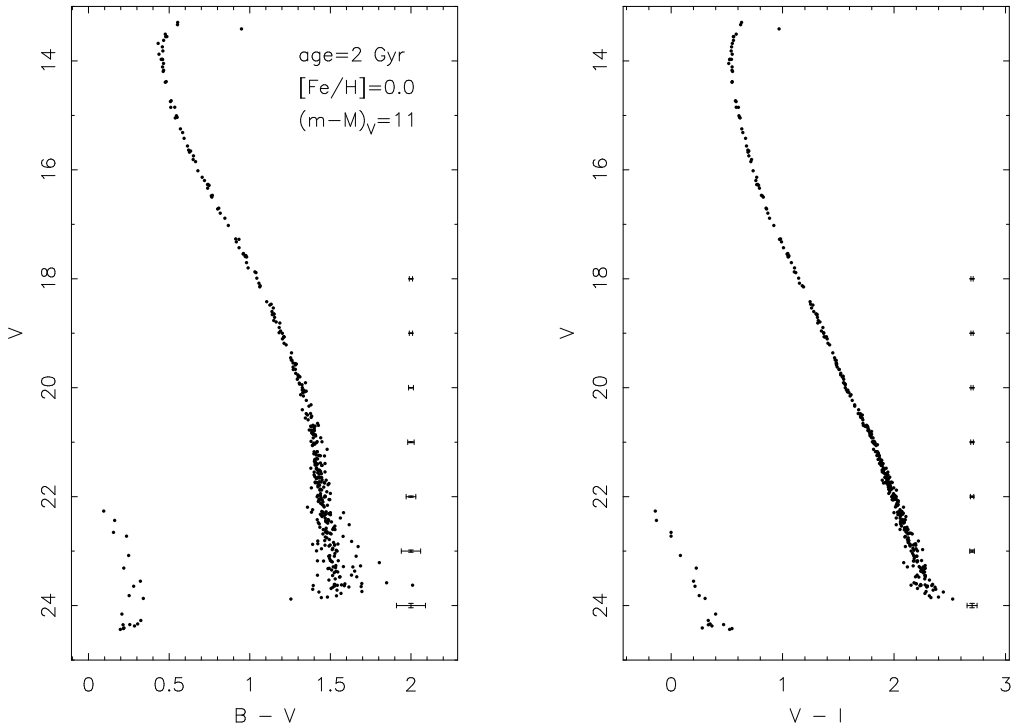}{3.7in}{270}{55}{55}{-200}{320}
\caption{$BV$ and $VI$ color-magnitude diagrams for an example simulated
cluster.  This cluster is 2 Gyr old, 1.5 kpc distant, has $A_{\rm V}$ = 0,
and contains 400 stars, 26 of which are WDs.  Photometric errors typical
for HST/ACS are included.  Average photometric errors as a function of
depth are presented in the error bars down the right hand side of each
color-magnitude diagram.} 
\end{figure}

In Figures 1--3 we present internal validation studies of our technique,
in which simulated clusters with different, but known, parameters are
presented to the Bayesian code for analysis.  In Figure 1 we present a
simulated 2 Gyr old star cluster as it might be observed with HST.  The
location of each star in the color-magnitude diagram is determined by
randomly drawing it from the IMF, assigning it the properties according to
the model ingredients discussed above, then scattering its photometry
based on the assumed errors as a function of depth.  The number of stars
and the errors are set to match those of a good target open cluster in one
or two HST/ACS fields.  We have not yet incorporated field stars, but this
will be done in the future.  For some clusters, field stars are not a
problem as they can be removed by proper motions and/or radial
velocities.  For other clusters this will not be possible, and our
technique will incorporate the probability of a star being a field star as
a function of its location in the color-magnitude diagram.

\begin{figure}[!ht]
\plotfiddle{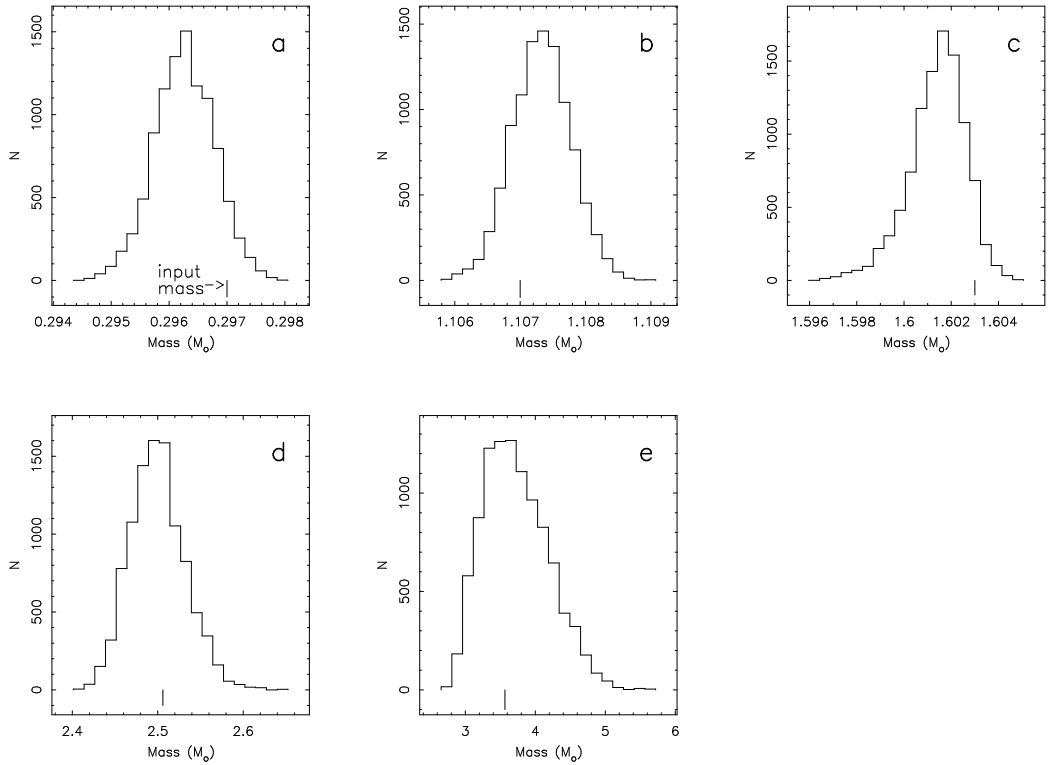}{3.82in}{270}{55}{55}{-200}{320}
\caption{Mass determinations for five example stars from the cluster of
Figure 1.  Panels {\it a} and {\it b} present main sequence stars, panel
{\it c} presents a red giant, and panels {\it d} and {\it e} present 
white dwarfs.  The input masses are marked with short vertical lines 
near the x-axes.} 
\end{figure}

Our Bayesian approach recovers the probability distributions for the
masses of every cluster star.  Five representative examples are presented
in Figure 2.  Note the very high precision (probability distributions with
widths of only a few thousandths of a solar mass) in the masses for the
main sequence and red giant stars.  These are internal measurements of
precision, of course, and do not represent the differences between
different model isochrones, i.e., these precisions do not represent the
external uncertainties in the mass-luminosity relation.  The white dwarf
masses, particularly as we plot them here, are much less precise, because
we plot their zero age main sequence mass and because a wide range in
initial masses becomes a narrow range in white dwarf mass.

\begin{figure}[!ht]
\plotfiddle{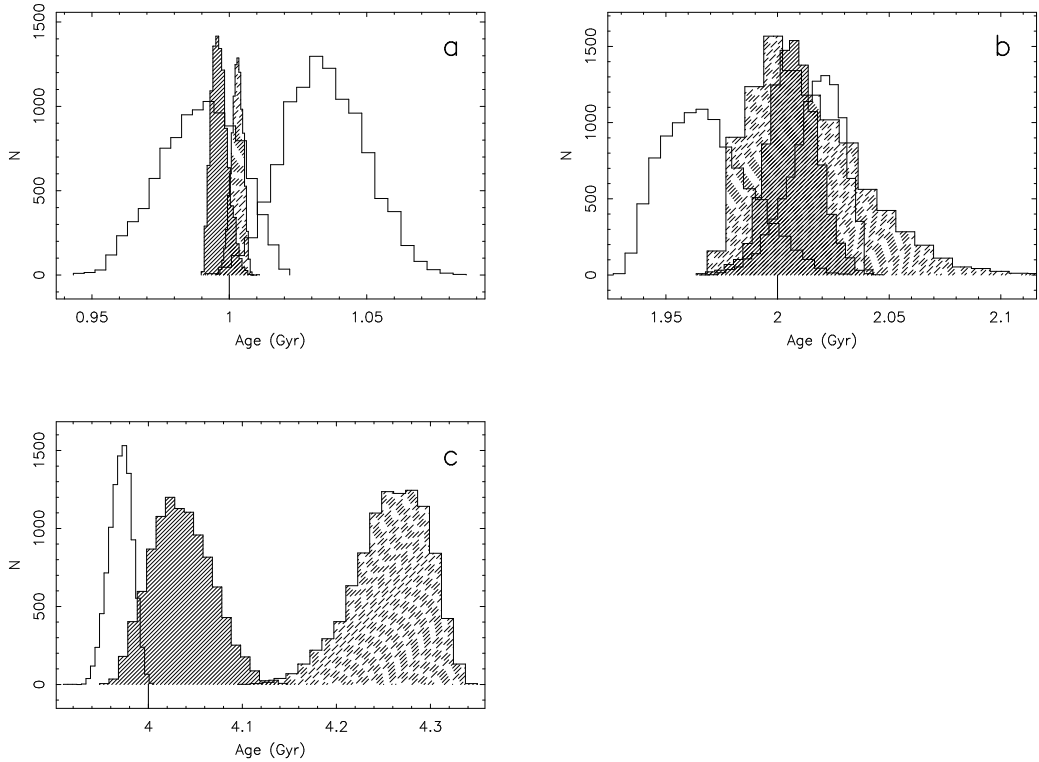}{3.8in}{270}{55}{55}{-200}{320}
\caption{Age determinations for eleven example clusters.  There are four
simulated clusters at 1 (panel {\it a}) and 2 Gyr (panel {\it b}), and
three simulated clusters at 4 Gyr (panel {\it c}).  The input ages are
marked with short vertical lines near the x-axes.  The remainder of the
cluster parameters are similar to those for the cluster presented in 
Figure 1.}
\end{figure}

The goal of our Bayesian analysis is to recover the probability
distributions for the cluster ages, and these are presented in Figure 3.
For this internal precision study, for the number of cluster stars
simulated (400), and for the assumed photometric error bars, the ages are
highly precise, with typical age distributions having a width of only
1--5\% of the cluster age.

\acknowledgements{We appreciatively acknowledge support for this research
from NASA through LTSA grant NAG5-13070.}

\vfill\eject

\end{document}